\documentclass[fleqn,usenatbib]{mnras}
\usepackage{newtxtext,newtxmath}
\usepackage[T1]{fontenc}
\usepackage{graphicx}	
\usepackage{amsmath}	
\usepackage{orcidlink}

\title[Disks around FFPMOs in IC348]{Disks around young free-floating planetary-mass objects:\\ Ultradeep Spitzer imaging of IC348}

\author[Seo \& Scholz]{
Holly Hanbee Seo,$^{\orcidlink{0009-0006-0039-6297} \ 1}$,
Aleks Scholz$^{\orcidlink{0000-0001-8993-50531} \ 1}$\thanks{E-mail: as110@st-andrews.ac.uk}
\\
$^{1}$SUPA, School of Physics \& Astronomy, University of St Andrews, North Haugh, St Andrews, KY169SS, United Kingdom\\
}

\date{Accepted 2025 January 22. Received 2025 January 10; in original form 2024 August 28}

\pubyear{\the\year{2024}}

\begin{document}
\label{firstpage}
\pagerange{\pageref{firstpage}--\pageref{lastpage}}
\maketitle

\begin{abstract}
Protoplanetary disks have been found around free-floating objects with masses comparable to those of giant planets. The frequency and properties of these disks around planetary-mass objects are still debated. Here we present ultradeep mid-infrared images for the young cluster IC348, obtained through stacking of time series images from Spitzer. We measure fluxes at 3.6 and 4.5$\,\mu$m for known free-floating planetary-mass objects (FFPMOs, spectral type M9 or later) in this cluster. By comparing the observed infrared spectral energy distributions with photospheric templates, we identify six planetary-mass objects with disks, plus three which may or may not have a disk. This corresponds to a disk fraction of $46\pm^{13}_{12}$\%. The disk fraction among planetary-mass objects is comparable to more massive brown dwarfs. We show the disk fraction among free-floating planetary-mass objects as a function of age, demonstrating that these objects retain disks for several million years, similar to low-mass stars and brown dwarfs.
\end{abstract}




\section{Introduction}

It is now common knowledge that our Galaxy hosts a large number of brown dwarfs (perhaps 100 billion altogether, \citet{muzic17}), as well as numerous isolated objects with masses around or below the Deuterium burning limit. First discovered in 2000 \citep{zapatero00,lucas00}, these free-floating planetary-mass objects have now been found in nearby young clusters, associations, as well as in microlensing surveys \citep{scholz12,clanton17,miretroig22}. The prevalence of FFPMOs is still under debate; as is their origin. Some of them could be the lowest mass objects to form like stars, others may be ejected planets. Future direct imaging and microlensing studies are expected to help us characterise the population of FFPMOs in the Galaxy \citep{scholz22,sumi23}.

The majority of stars, as well as most brown dwarfs, start their lives with a circum-(sub)-stellar accretion disk that lasts for one or more million years. In many cases planets will form from the raw material in the disks, giving rise to the rich and diverse population of planetary systems established by exoplanet surveys \citep{cassan12}. The early signs of planet formation in the disks -- in the form of grain growth and gap formation -- have been found down to central object masses below 0.1$\,M_{\odot}$ \citep{scholz07,pinilla18}. Moreover, the disks themselves have been identified around objects with masses as low as 0.01$\,M_{\odot}$, or $\sim 10\,M_{\mathrm{Jup}}$ \citep{luhman05b,joergens13,scholz23,damian23}. The standard way to identify disks is by looking for infrared excess emission at wavelengths $>2\,\mu m$, an approach that has benefitted from the wide availability of sensitive coverage of star forming regions by the Spitzer Space Telescope \citep{werner04}.

Here we investigate the prevalence and lifetimes of disks around young FFPMOs, using ultradeep images created by stacking archival Spitzer data. This paper is a direct follow-up to a previously published study with similar approach and goals, which was focused on the young cluster NGC1333 \citep{scholz23}. Our target for the current paper is IC348, the second major cluster in the Perseus star forming complex \citep{luhman16,pavlidou21}. With nearly 500 cluster members and a quoted age of 3-5\,Myr, IC348 is larger and slightly older than its sibling NGC1333. As in \citet{scholz23}, we use Spitzer/IRAC time-series imaging to produce ultradeep images of the cluster, and measure infrared fluxes for substellar members (Section \ref{sec:data}). We identify planetary-mass objects with disks, derive the disk fraction, and compare with the literature (Section \ref{sec:disks}). We present our conclusions in Section \ref{sec:sum}.

\section{Data processing}
\label{sec:data}

\subsection{Stacking of Spitzer images}
\label{sec:img_stack}

The cluster IC348 was observed by Spitzer several times. Here we make use of a time series obtained during the 'warm mission', in cycle 9, as part of the program with the ID 60160 (PI: Muzerolle). Altogether, 38 images were taken in the IRAC1 and IRAC2 channels, at wavelengths of 3.6 and 4.5$\,\mu m$ \citep{fazio04}, and at a constant position. The observations are spread out over 40 days in late 2009. The primary scientific purpose of this program was to study variability, with the results reported in \citet{flaherty11,flaherty12}. Here we stack all data for a given channel to produce an ultradeep image of the cluster. The single-frame exposure time is 12\,sec, identical to single-epoch mosaics obtained for IC348 for the 'Cores to Disk' program \citep{evans09}. that means, the stack will have a combined on-source time of 456\,sec (or 7.6\,min).

We downloaded all 38 images per channel from the Spitzer Heritage Archive. These two sets were stacked using the Python package {\tt reproject}, a part of {\tt astropy}. The full stacked image covers approximately $0.4 \times 0.4$ degrees, including the majority of the known cluster members. The central J2000.0 coordinates in IRAC1 are (56.08,32.10) and (56.12,32.00) in IRAC2. The pixel scale of the stacks, as well as the individual images, is 0.6\,arcsec. By stacking 38 images, the signal-to-noise ratio should be improved by about $\sqrt{38} = 6.16$, which means the resulting image should be about 2\,mag deeper than an individual epoch.
 
\subsection{Brown dwarf sample}

To define our sample, we use the census for this cluster from \citet{luhman16}, which encompasses in total 478 stars and brown dwarfs classified as members of IC348. According to the authors, the survey is 'nearly complete' down to K-band magnitudes of 16.8 and J-band extinction $A_J=1.5$. For our purposes, we selected the subsample of objects with spectral types of M6 or later. According to evolutionary tracks, this spectral type corresponds to a mass around 0.1$\,M_{\odot}$ for ages of 1-5\,Myr; hence, our subset should include all spectroscopically confirmed brown dwarfs in this census. In this paper, however, we are primarily interested in the sources with spectral types M9 or later, which is likely to correspond to masses around or below the Deuterium burning limit at young ages.

\begin{figure}
\includegraphics[width=\columnwidth]{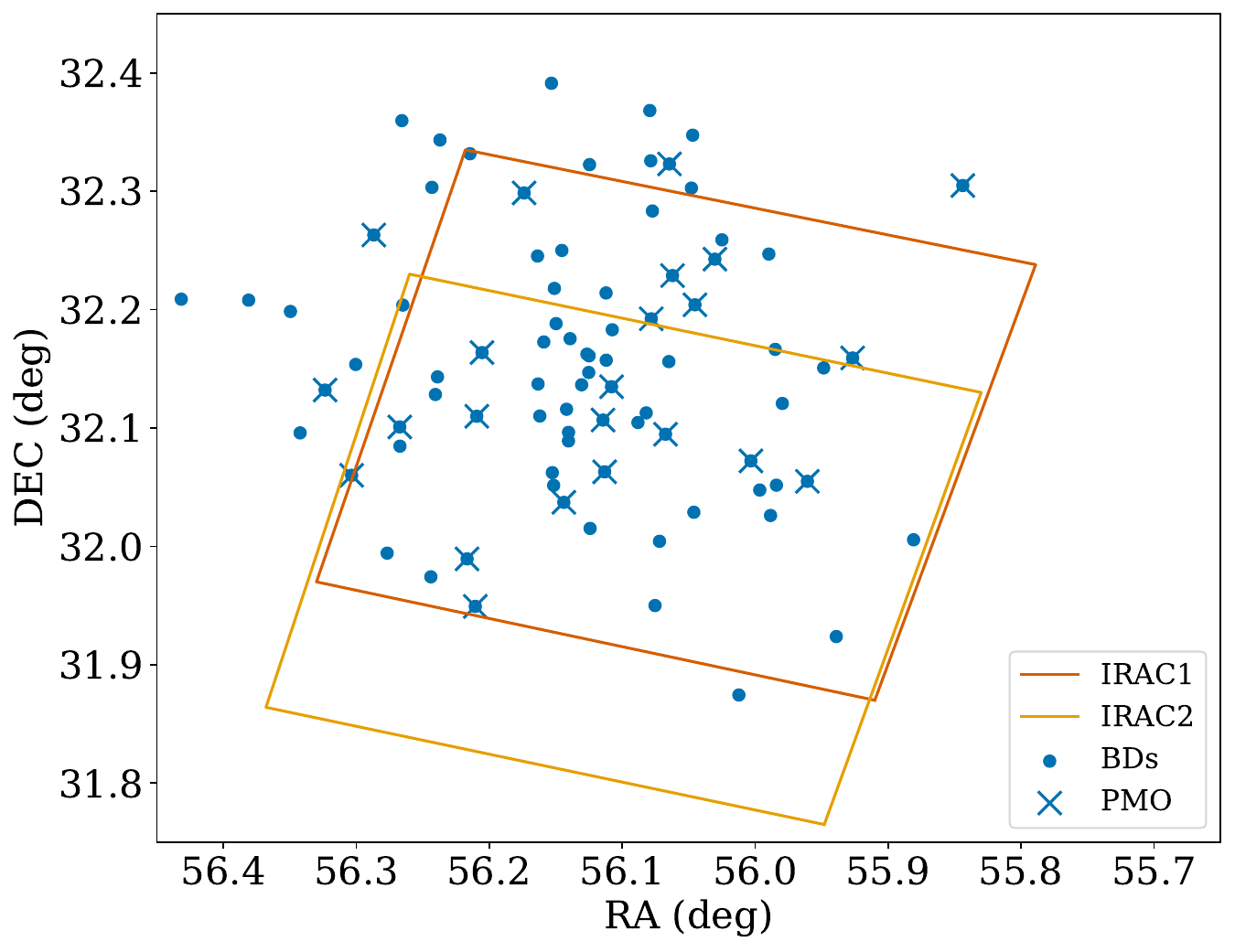} 
\caption{Spatial distribution of brown dwarfs in IC348, defined as spectral type M6 or later. In red and orange, we show the approximate coverage of the fields in IRAC1 and IRAC2 for the stacked image used in this paper. We mark the planetary-mass objects with spectral type M9 or later.}
\label{fig:field}
\end{figure}

In total, we select 83 objects with adopted spectral type of M6.0 or later. Of these, 23 have spectral type of M9 or later. From this subsample, 67 are covered by our deep Spitzer images, 19 of which have spectral type of M9 or later. Due to the offset between the two IRAC channels, some objects are only covered in one band. In addition, some objects are located too close to the edges for reliable measurements, those will not be used in the following. Overall, this results in 17 objects with M9 or later spectral type and valid measurements. Four of those are only covered in IRAC1. In Figure \ref{fig:field} we show the spatial distribution of the sample, compared to the fields covered by our deep IRAC1 and IRAC2 image. We note that a recent JWST survey uncovered a couple more objects which may be planetary-mass members of IC348 \citep{luhman24}, but are too faint to be included here.

\subsection{Photometry}

We measured fluxes for the brown dwarfs in the images using {\tt photutils}. For the circular apertures, we used a radius of 5 pixels (or 3 arcsec). The background was estimated as the median in a circular annulus between radii of 7 and 9 pixels. After converting to magnitudes, those were calibrated by cross-matching with the catalogue by \citet{gutermuth09} which include the IC348 region.

To calculate extinction-free colours, we dereddened the IRAC1 and IRAC2 photometry, using the extinction measurements provided by \citet{luhman16}. Their paper lists $A_J$, calculated from $A_V$ with the \citet{cardelli89} extinction law ($A_J = 0.28A_V$). We re-derived $A_J$ using the more recent extinction law from \citet{wang19} ($A_J = 0.243 A_V$). We used the same extinction law to deredden the photometry in the IRAC bands. We dereddened the published K-band photometry by \citet{luhman16} using the same process. 

The median photometric error for our sample is 0.07\,mag. For faint objects, the photometric error can be significantly larger; for the planetary-mass objects in our sample the error in the IRAC photometry is up to 0.2\,mag. Ad additional error from the calibration we adopt 0.01\,mag. Also, the extinction correction adds a small uncertainty, which combines the error in $A_V$ ($\sim 0.5$\,mag) and the negligible error in the band coefficient (0.003-0.004\,mag). The magnitudes and dereddened colours for all planetary-mass objects in our sample are listed in Table \ref{tab:phot}.

\begin{table*}
\caption{New Spitzer photometry for planetary-mass objects in IC348. Spectral type, $A_J$, $K$ are from \citet{luhman16}, but the extinction has been adjusted with the more recent extinction law by \citet{wang19}. The colours are dereddened to $A_J=0$, and are identical to the values plotted in Figure \ref{fig:kirac}.}
\label{tab:phot}
\begin{tabular*}{0.8\textwidth}{lllcccccc}
\hline
UGCS                & LRL   & SpT   & $A_J$ & K    & I1    & I2    & $K-IRAC1$ & $K-IRAC2$ \\
\hline
J034350.58+320317.5 & 1843  & M9    & 2.17 & 15.19 & 13.89$\pm0.06$ & 13.34$\pm0.06$ & 0.94$\pm0.09$ & 1.38$\pm0.09$ \\ 
J034416.18+320541.0 & 4044  & M9    & 0.47 & 17.09 & 15.20$\pm0.11$ & 14.75$\pm0.11$ & 0.81$\pm0.13$ & 1.24$\pm0.13$ \\ 
J034427.18+320346.6 & 705   & M9    & 0.35 & 15.73 & 14.99$\pm0.10$ & 14.83$\pm0.12$ & 0.68$\pm0.12$ & 0.83$\pm0.13$ \\ 
J034504.17+320602.9 &       & M9    & 0.00 & 16.87 & 16.04$\pm0.17$ & 15.97$\pm0.20$ & 0.83$\pm0.18$ & 0.90$\pm0.21$ \\ 
J034449.33+320949.4 &       & M9    & 0.10 & 16.56 & 15.82$\pm0.15$ & 15.74$\pm0.18$ & 0.72$\pm0.16$ & 0.80$\pm0.19$ \\ 
J034449.33+320949.4 & 40023 & M9.5  & 0.39 & 16.88 & 15.76$\pm0.15$ & 15.56$\pm0.17$ & 1.05$\pm0.16$ & 1.24$\pm0.18$ \\ 
J034450.24+320635.5 &  5231 & M9.5  & 0.00 & 17.39 & 16.65$\pm0.22$ & 16.06$\pm0.21$ & 0.74$\pm0.23$ & 1.33$\pm0.22$ \\  
J034451.99+315921.6 & 1379  & M9.7  & 0.29 & 15.86 & 14.95$\pm0.10$ & 14.57$\pm0.11$ & 0.86$\pm0.12$ & 1.23$\pm0.12$ \\  
J034400.75+320420.0 & 40142 & L0    & 0.00 & 17.69 & 16.76$\pm0.23$ & 16.51$\pm0.26$ & 0.93$\pm0.24$ & 1.18$\pm0.27$ \\ 
J034450.52+315657.3 &       & L0    & 0.00 & 16.39 & 15.68$\pm0.14$ & 15.55$\pm0.17$ & 0.71$\pm0.16$ & 0.84$\pm0.18$ \\ 
J034427.47+320624.2 & 30057 & L1    & 0.49 & 18.06 & 16.72$\pm0.23$ & 16.17$\pm0.22$ & 1.23$\pm0.24$ & 1.79$\pm0.23$ \\  
J034425.92+320805.4 & 6005  & L1    & 0.49 & 17.54 & 16.66$\pm0.22$ & 16.17$\pm0.22$ & 0.80$\pm0.23$ & 1.26$\pm0.23$ \\ 
J034434.54+320213.7 & 5209  & L1    & 0.49 & 17.48 & 16.63$\pm0.22$ & 16.31$\pm0.24$ & 0.77$\pm0.23$ & 1.06$\pm0.25$ \\ 
\hline
\end{tabular*}
\end{table*}

\section{Disks in planetary-mass objects}
\label{sec:disks}

\subsection{Infrared colour excess}

Circum-sub-stellar disks are routinely identified by looking for infrared colour excess caused by warm dust in the inner parts of the disks. Here we use the $K-IRAC$ colour (combining our IRAC photometry with the K-band magnitudes from \citet{luhman16} as an initial indicator for the presence of a disk. For brown dwarfs and planetary-mass objects with temperature below 3000\,K, the $K$ band magnitudes at a wavelength of 2.2$\,\mu m$ are dominated by the photosphere. On the other hand, the IRAC magnitudes at wavelengths of 3.6 and 4.5$\,\mu m$ are affected by excess emission from the disk, if present \citep[e.g.][]{natta01}. Overall, the $K-IRAC1$ and $K-IRAC2$ colours should show excess above the photosphere, if a disk is present.

\begin{figure}
\includegraphics[width=\columnwidth]{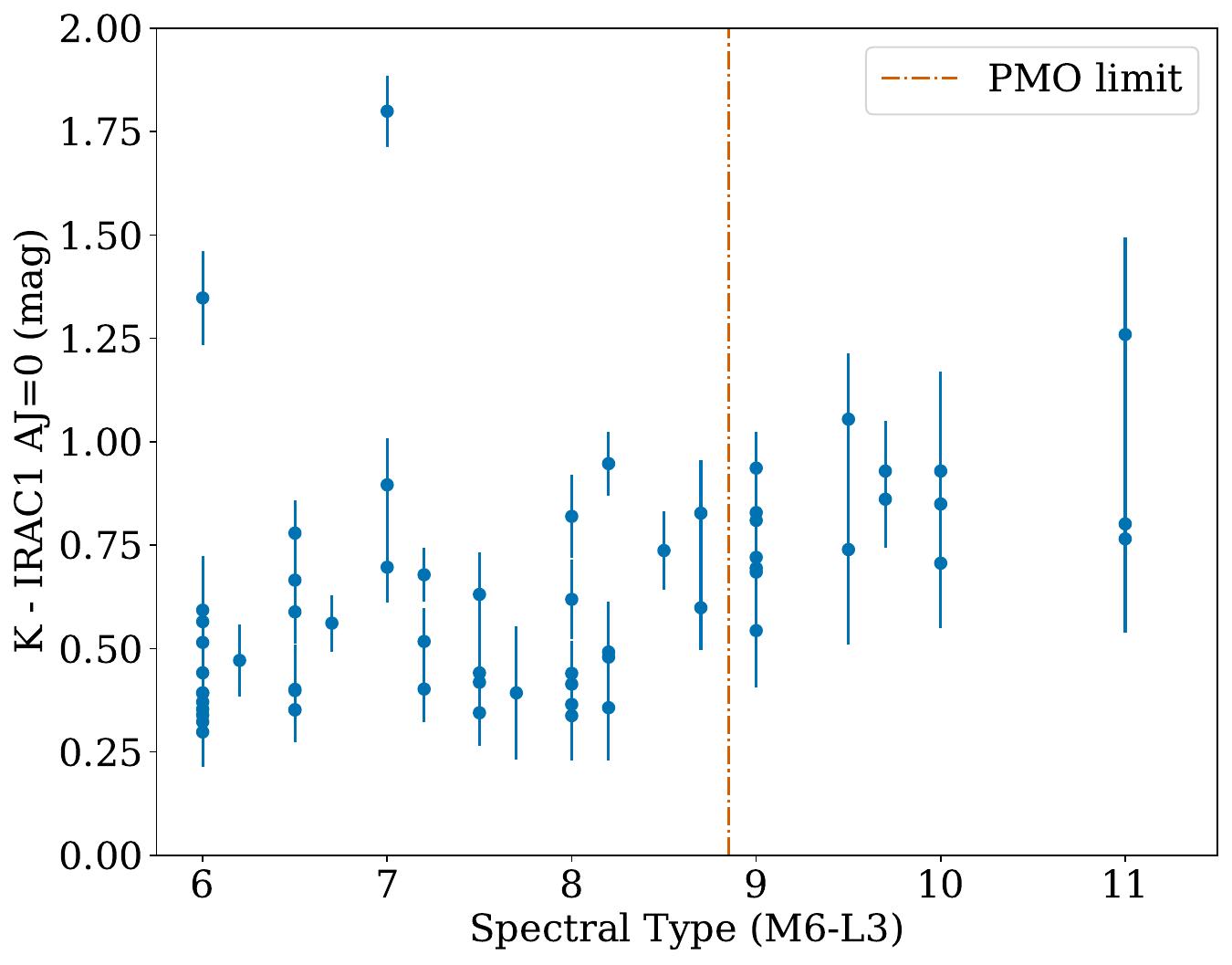}
\includegraphics[width=\columnwidth]{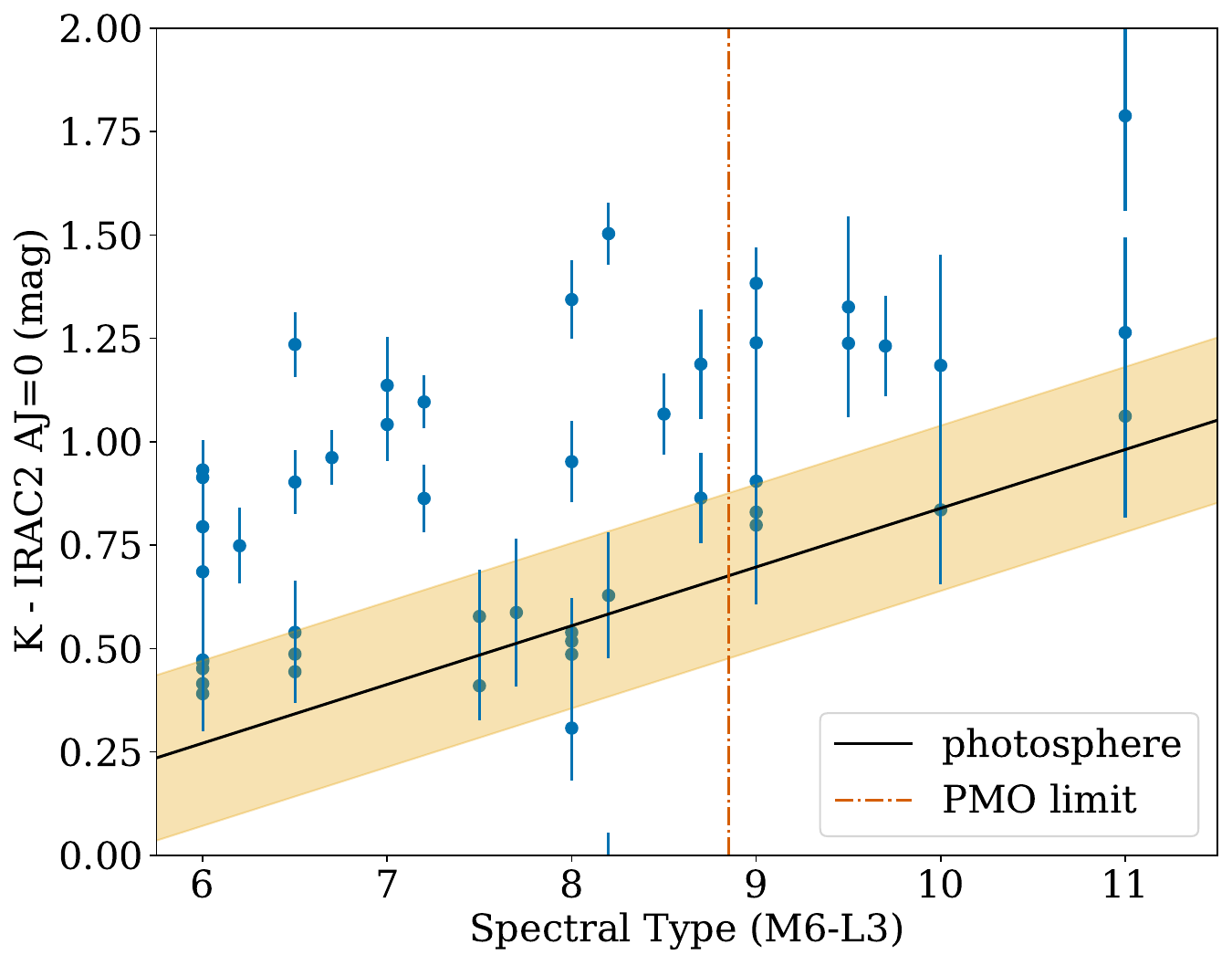}
\caption{$K-IRAC$ infrared colours vs spectral type for all objects with valid measurements. In this paper we primarily focus on the objects with spectral types M9 or later, on the right side of the red dash-dotted line. For $K-IRAC2$ the separation between objects with/without disks is apparent; for this panel we also overplot the photospheric colours. For more details, see text.}
\label{fig:kirac}
\end{figure}

In Figure \ref{fig:kirac}, we show the dereddened $K-IRAC2$  and $K-IRAC1$ colours as a function of spectral for our sample. In $K-IRAC2$ we see two populations of objects forming two sequences. The population with elevated colour should correspond to objects with disks. In $K-IRAC1$ the division is not clear, and it is difficult to disentangle objects with and without excess. This is expected, the signature of the disks should become more distinctive at longer wavelengths. In the following, we will therefore use exclusively the $K-IRAC2$ colour.

To robustly identify a colour excess requires an estimate of the photospheric colour. We use the polynomial relations from \citet{sanghi23} published in their table A1, for this purpose. They provide relations for absolute magnitude vs. spectral type for brown dwarfs in young moving groups. Instead of the IRAC bands, they present relations for the WISE bands at similar wavelengths (W1 and W2). We compared the IRAC with WISE photometry for young, cool stars and brown dwarfs from a sample in the Upper Scorpius star forming region \citep{esplin18}. This region is slightly older than IC348 and the overwhelming majority of the objects do not show excess emission. The $K-IRAC2$ colour in that sample follows closely the $K-W2$ colour, confirming that using the \citet{sanghi23} relations for WISE is valid for estimating the photospheric colours. We note that this relation is also comparable to the one adopted in \citet{scholz23}, which was based on an independent study but does not extend beyond spectral types L1.

As a sidenote, using the same approach we find a significant offset of $\sim 0.2$\,mag between the $K-W1$ and the $K-IRAC1$ colours, The W1 and IRAC1 bands are similar \citep{jarrett11}, but not identical. Late M to early L dwarfs feature broad absorption due to water in this band \citep{cushing05}, which means small changes in the band limits can affect the colours. Conversely, in the IRAC2 band the spectrum for late M/early L objects is essentially a smooth slope \citep{manjavacas24}, which simplifies band conversions.

The photospheric values are shown as black line in the lower panel of Figure \ref{fig:kirac}, with an indicative uncertainty as yellow band. The plot clearly demonstrates that some of the sources show colour excess above the photosphere, i.e. their colours are located above the range expected for a pure photosphere. To re-iterate, all objects in these plot were checked individually in the images, to ensure that close neighbours, nearby edges, and varying background do not affect the flux measurements. Judged by the $K-IRAC2$ colour, the fraction of sources with infrared excess and spectral type of M9 or later is a maximum of 9 of 13, with considerable uncertainty. In the following we scrutinise these 9 potential PMOs with disks more in detail. For the brown dwarfs with M6-M8 spectral type, the disk fraction would be around 50\% according to this figure, with a few objects that are difficult to classify. As we will discuss later, this disk fraction is in line with previous estimates for the mid to late M dwarfs in this cluster. 

\subsection{Spectral energy distributions}

To verify the presence of disks in the sources with spectral types M9 or later, we combine our photometry at 3.6 and 4.5$\,\mu m$ with literature data. In particular, we use the deep J, H, and K photometry from UKIDSS-DR9 \citep{lawrence07,lawrence13}, which is available for all objects in our sample. For the very faintest source, we use the J-band magnitude from \citet{preibisch03}. Whenever we found multiple valid measurements for the same band, we verified that they are in agreement with each other. We de-redden the UKIDSS and our IRAC magnitudes, using the values of $A_J$ used earlier and the \citet{wang19} extinction law, and convert to flux densities $F_{\nu}$ in mJy. In addition, if available, we add the Spitzer fluxes at 5.8 and 8.0$\,\mu m$ from the C2D Full Clouds Catalogue \citep{evans09}. In general, if a flux has been measured by C2D at 5-8$\,\mu m$, it is a clear confirmation of a disk. The photospheric fluxes for objects in IC348 at those spectral types are too low beyond 5$\,\mu m$ to be detected robustly by Spitzer. To unambiguously identify a disk we demand that the excess is seen in at least two bands.

As a template for the photospheric emission, we use here model spectra from the BT-Settl-AGSS series \citep{allard07,asplund09}, for $\log{g} = 3.5$ and effective temperatures ranging from 1500 to 2500. According to \citet{sanghi23} (see their Figure 16), this temperature range should cover the spectral type range considered here. Models are smoothed and scaled to the dereddened near-infrared fluxes of our targets. We varied the temperature until we achieve a good match between the model and the fluxes in J- and H-band which would represent the photosphere. The outcomes described below do not depend on the specific choice of model spectra within those parameters. 

\begin{figure*}
\includegraphics[width=0.65\columnwidth]{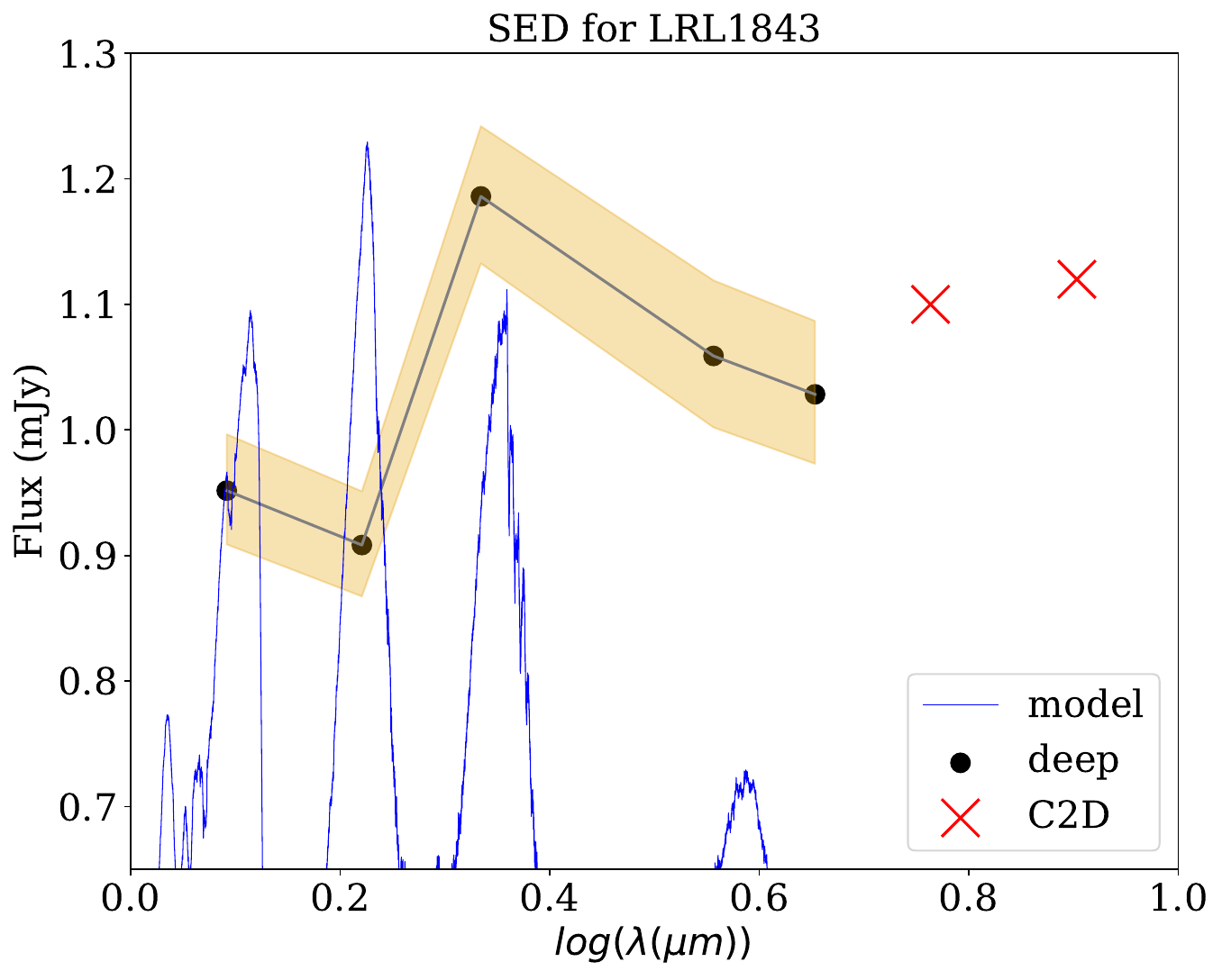}
\includegraphics[width=0.65\columnwidth]{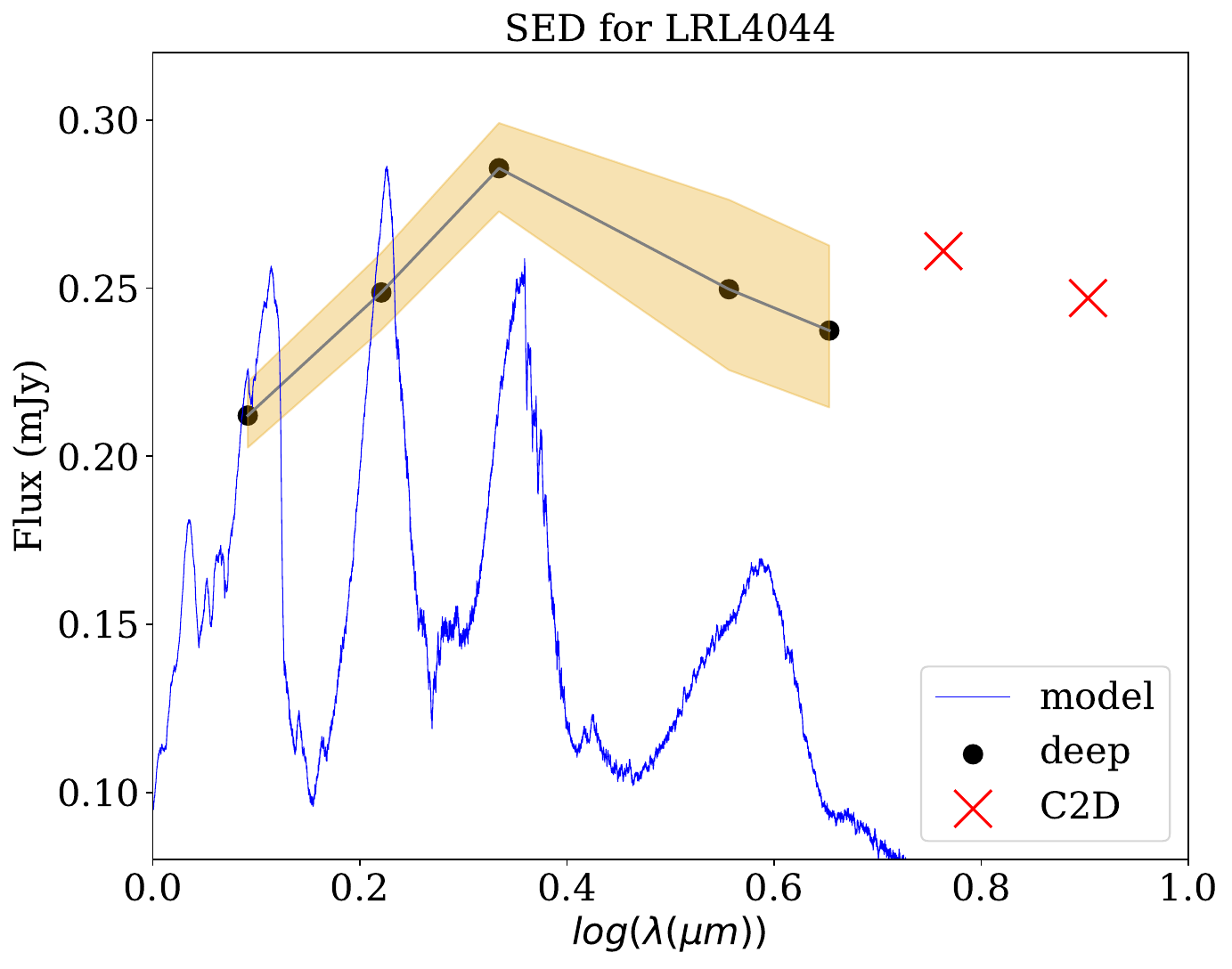}
\includegraphics[width=0.65\columnwidth]{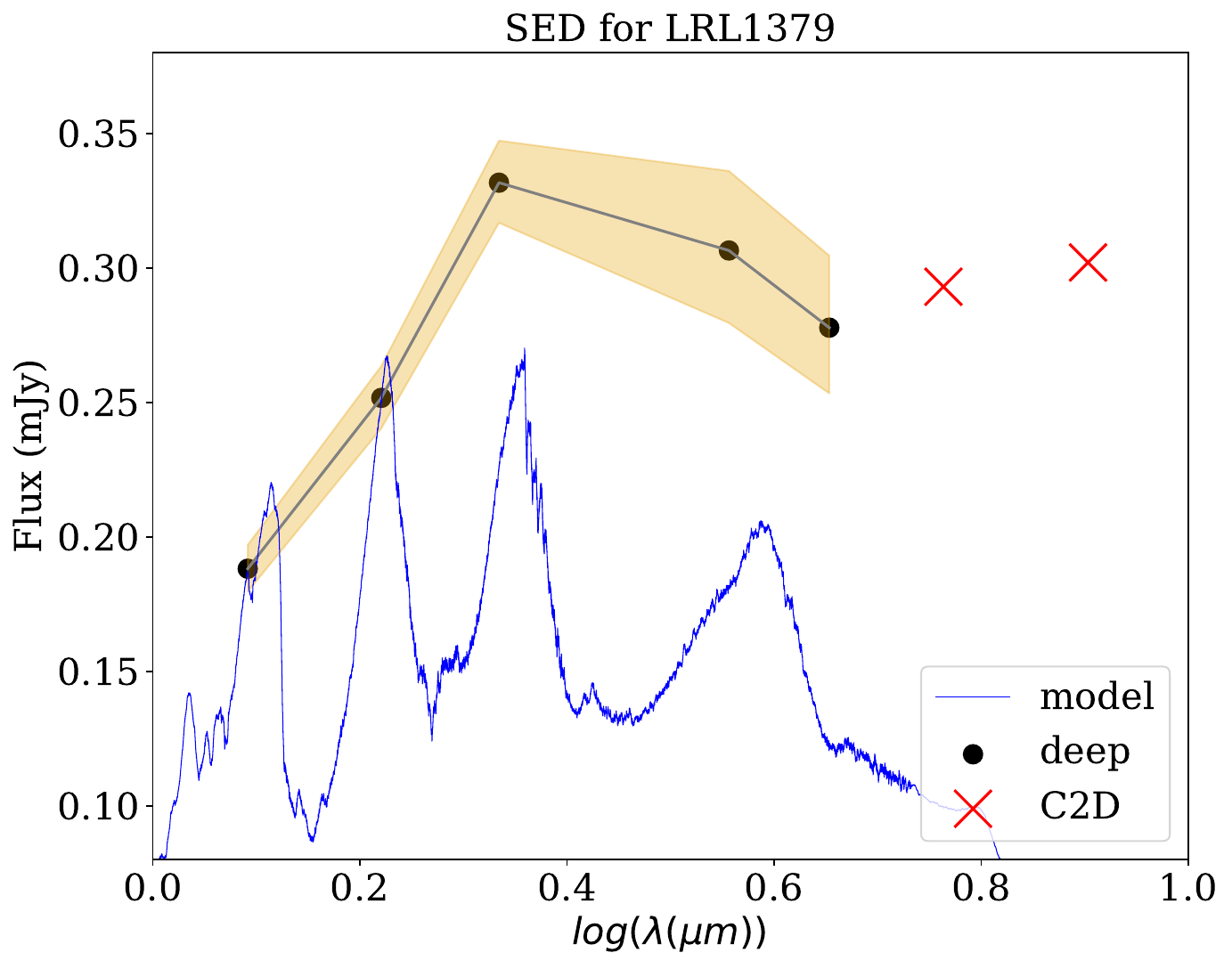}\\
\includegraphics[width=0.65\columnwidth]{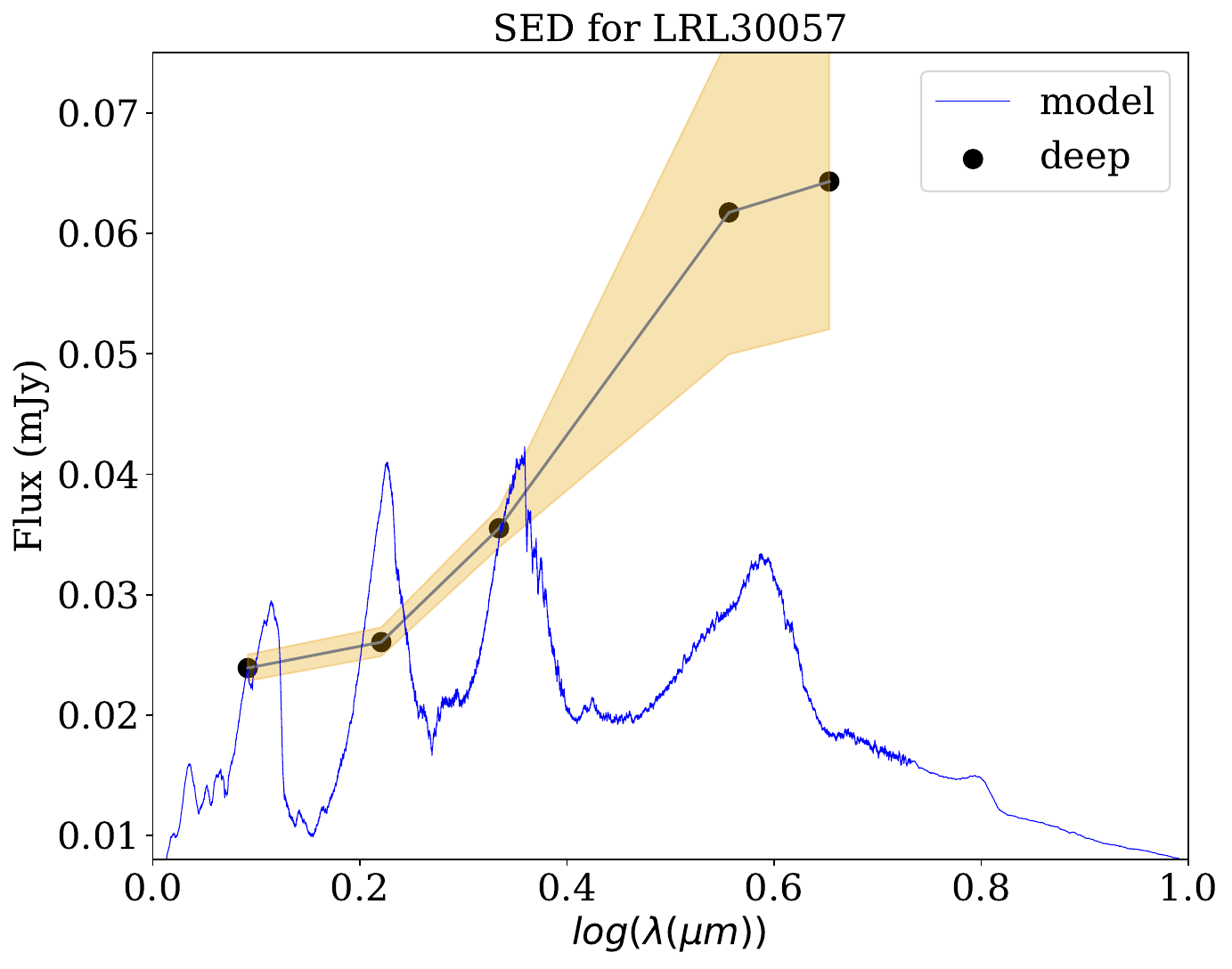}
\includegraphics[width=0.65\columnwidth]{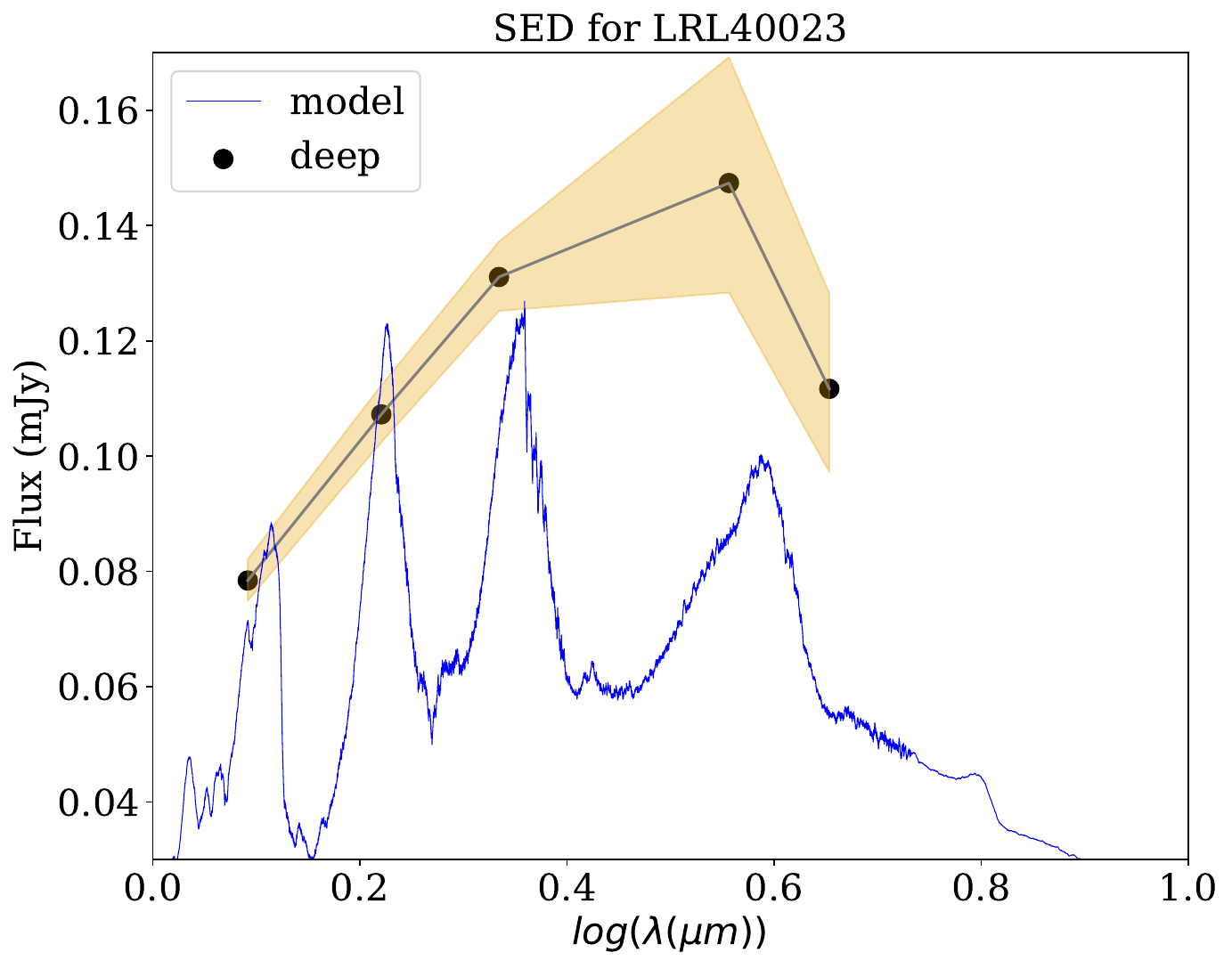}
\includegraphics[width=0.65\columnwidth]{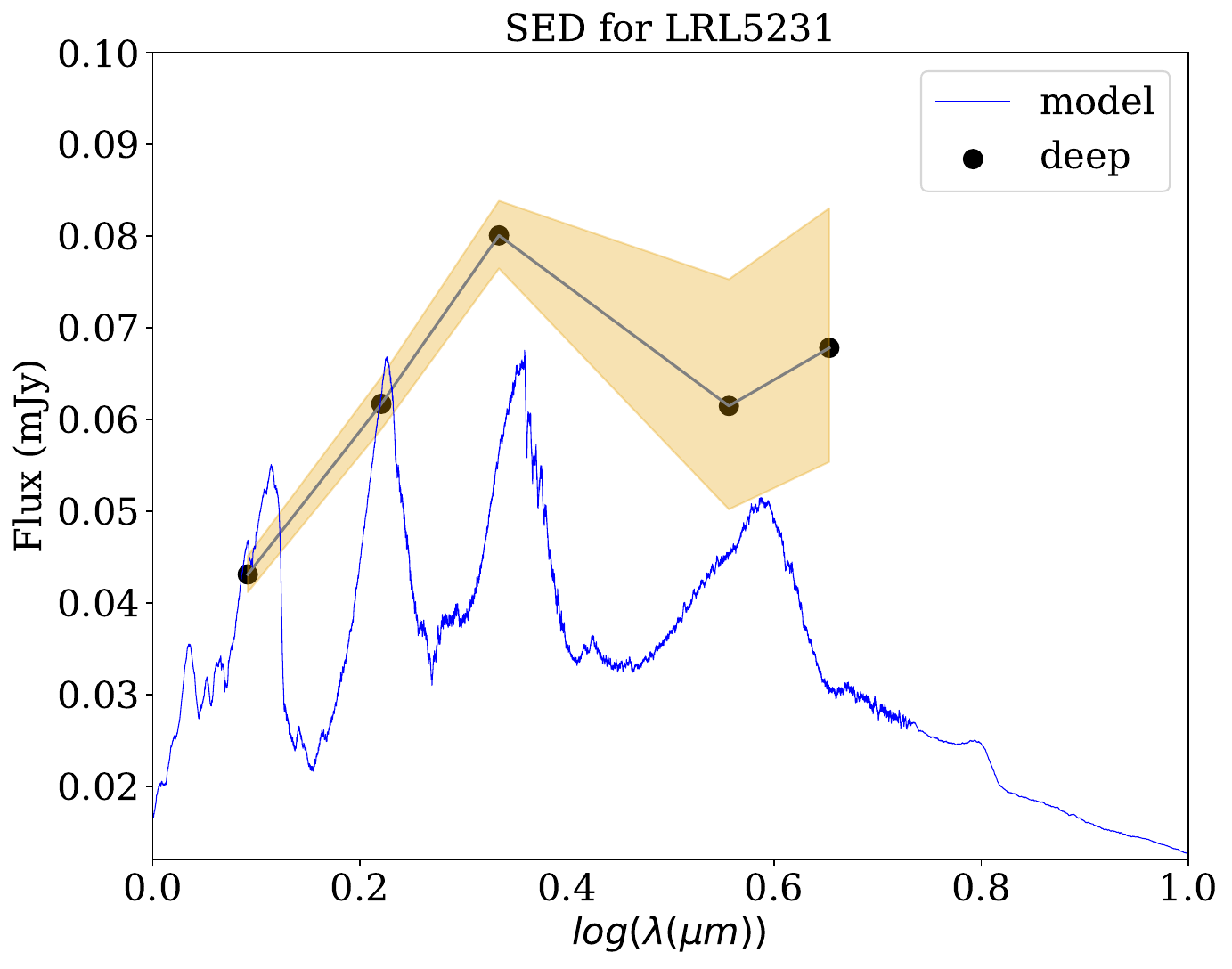}\\
\includegraphics[width=0.65\columnwidth]{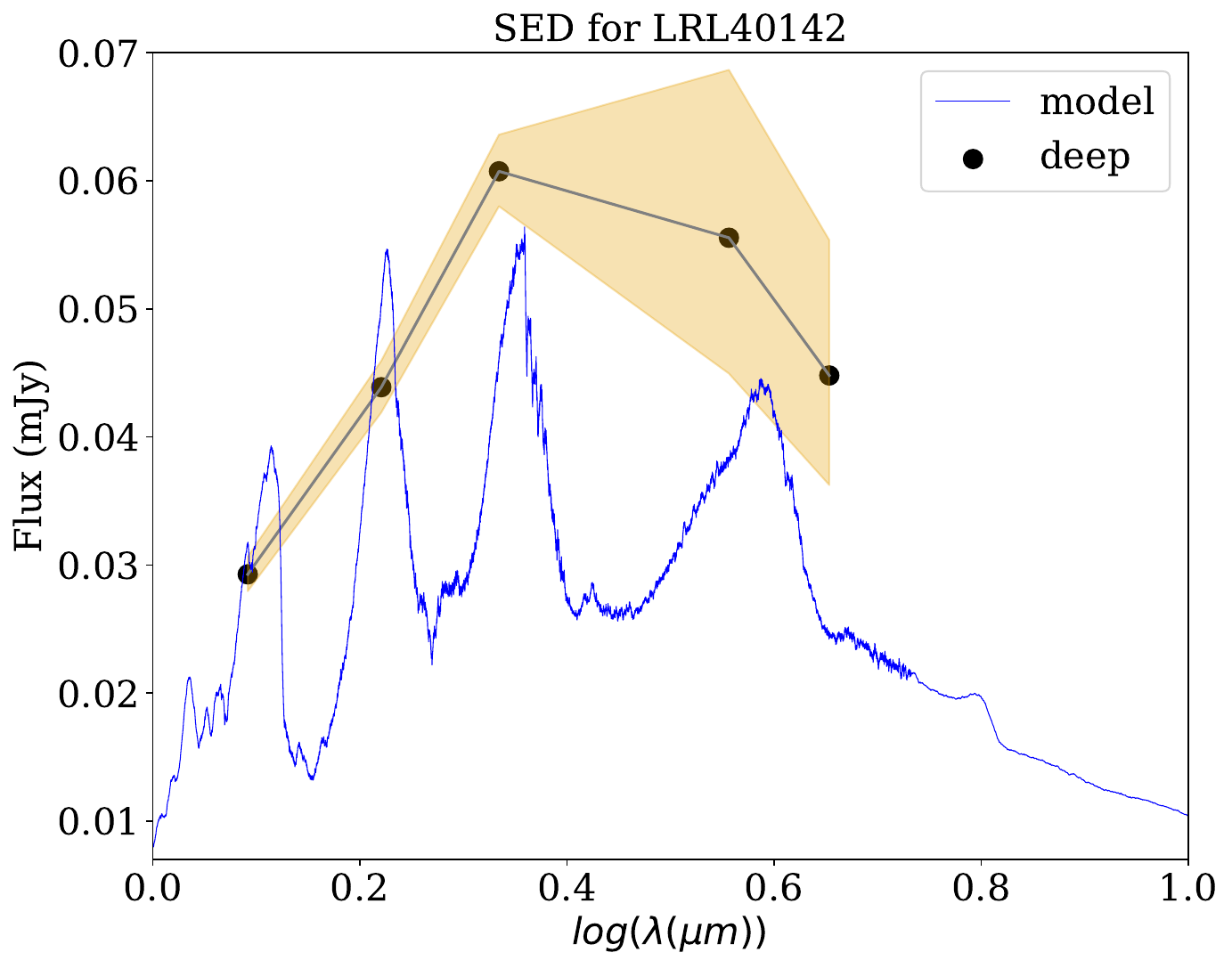}
\includegraphics[width=0.65\columnwidth]{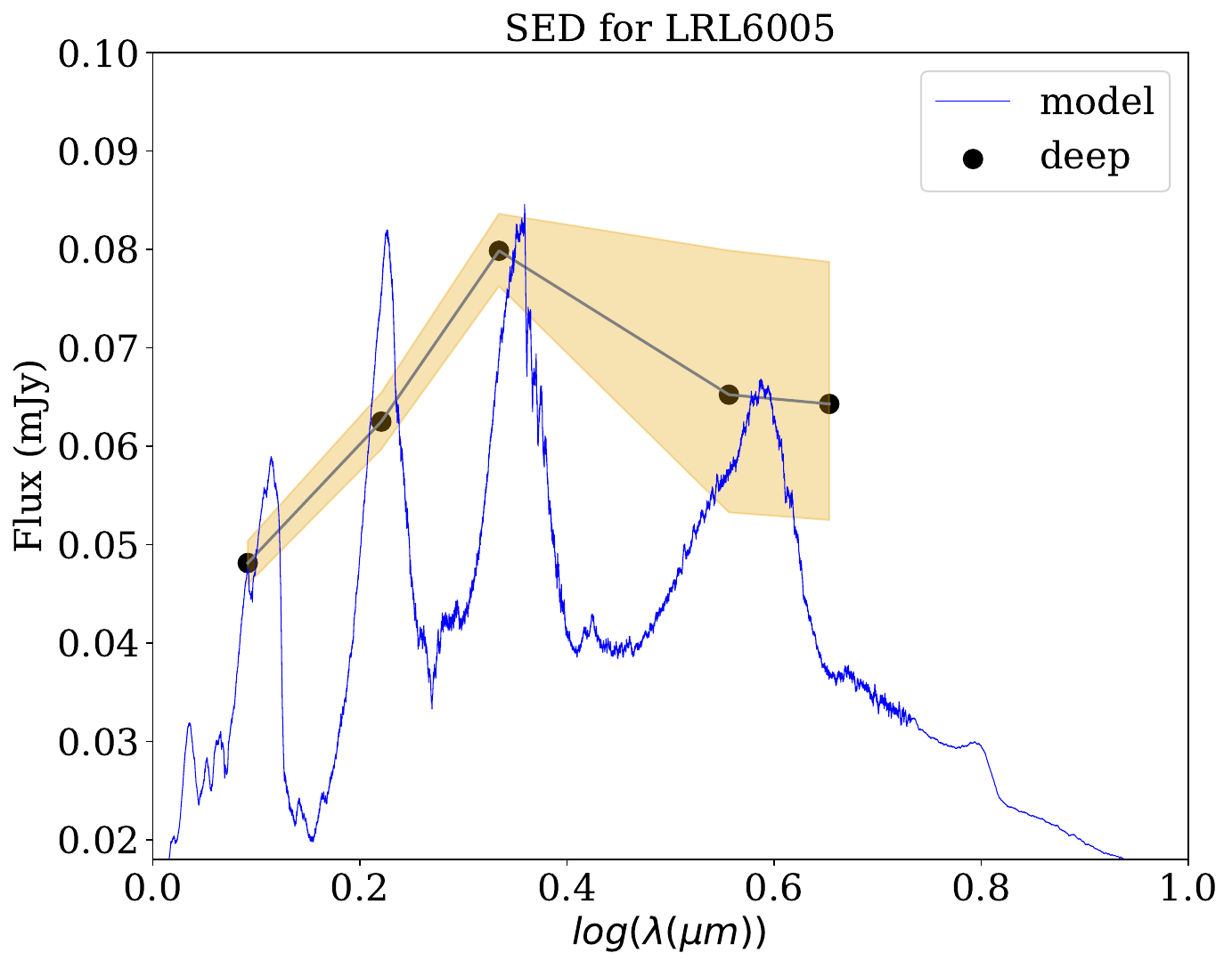}
\includegraphics[width=0.65\columnwidth]{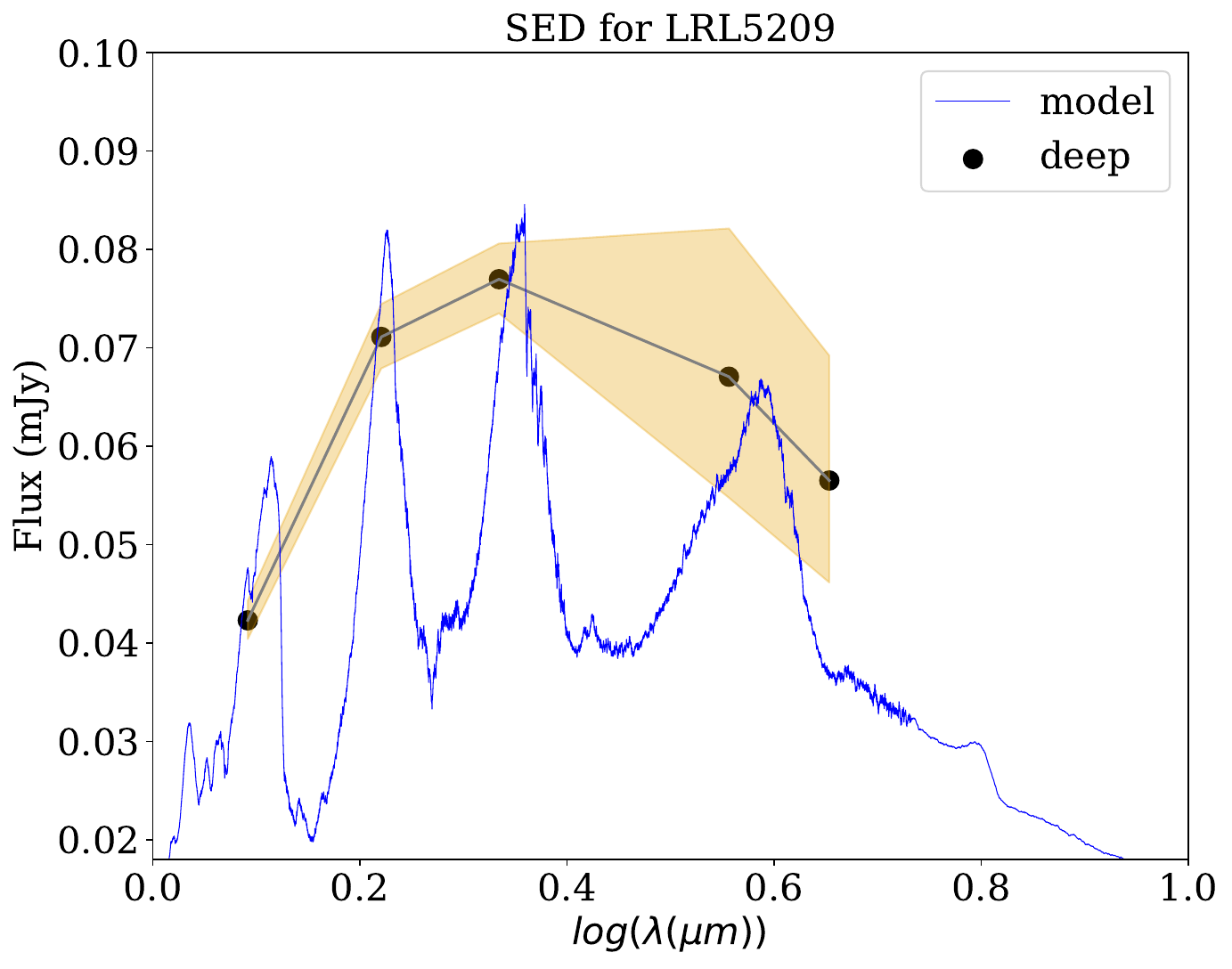}\\
\caption{Spectral energy distributions for planetary-mass brown dwarfs with potential excess emission due to disks. The black, connected datapoints are based on dereddened photometry, with the datapoints at 3.6 and 4.5$\,\mu m$ from this current study. The yellow shaded region indicates the error; here we adopt a conservative error of 5\% for the JHK photometry. The spectrum plotted in blue comes from models used as templates. When available, we also show the C2D datapoints at 5.8 and 8.0$\,\mu m$ in red. See text and Table \ref{tab:pmo} for details.}
\label{fig:sed}
\end{figure*}

In Figures \ref{fig:sed} we show the spectral energy distributions for all planetary-mass objects with potential disk detection, with errors as shaded region. In Table \ref{tab:pmo} we report the data used for these figures, plus the $T_{\mathrm{eff}}$ value of the model chosen as comparison. Based on this analysis, 5 of these 9 objects are unambiguously classified as having infrared excess, evidence for a circum-sub-stellar disk. For all 5, there are multiple measurements significantly above the photospheric flux. For 3 of these, at least one datapoint at 5-8.0$\,\mu m$ exists and confirms the excess. For one more there is marginal excess emission and an increasing SED; we add this to the sample with disks. For the remaining three, the figures do not show evidence for excess emission. The other 4 objects with spectral type of M9 or later in our deep images also have colours consistent with a pure photosphere. To re-iterate, this does not preclude the presence of a disk, they could still harbour a disk with an inner hole and excess at wavelengths longer than 8$\,\mu m$.

\subsection{Comments on individual objects}

In the following we provide some commentary on individual sources. 

LRL1843 has clear excess flux in the IRAC bands, compared to all templates with 2000 to 2500\,K. The excess emission also exceeds the (small) error in the infrared fluxes. The presence of the disk is also confirmed by robust detections at 5.8 and 8.0$\,\mu m$, with fluxes around 1\,mJy. Infrared excess has also been identified by \citet{luhman16} for this object. It is a clear detection of a disk. Note that the near-infrared fluxes do not match very well with the model spectra; this may be due to an overestimate in the extinction (which is the highest in this sample.)

LRL4044 shows clear excess in our IRAC measurements above the templates with plausible temperature. It also has a detection at 5.8$\,\mu m$ in C2D (but only an upper limit at 8.0$\,\mu m$). The SED is flat out to 6$\,\mu m$, whereas the photospheric template fluxes drop off quickly in the mid-infrared. This is a clear disk detection.

LRL1379 shows significant excess emission from 3 to 8$\,\mu m$, including robust detections in the C2D survey. There might already be excess emission in the K-band. The infrared excess was already identified in \citep{luhman16}. The object is clearly harbouring a circum-sub-stellar disk.

LRL30057 has a rising SED from 1 to 4.5$\,\mu m$, and substantial excess emission over photospheric templates at temperatures of 1900\,K or similar. The SED shows all characteristics of a Class II source with a flared disk. It is not robustly detected at longer wavelengths, but this is not surprising given how faint the source is. This object is regarded a safe disk detection.

LRL40023 has an SED that peaks at 3.6$\,\mu m$, in terms of $F_\nu$. It shows significant excess emission in IRAC1 and IRAC2, but has no robust C2D detection at longer wavelengths. In agreement with \citet{luhman16}, we still consider this object to be a safe disk detection, despite the decline in the flux level at 4.5$\,\mu m$.

LRL5231 has marginal excess emission in IRAC1, and substantial excess in IRAC2, compared to the 2000\,K template. There may be slight excess emission in the K-band as well. The flux at 4.5$\,\mu m$ is comparable or higher than at $3.6\,\mu m$, while the photospheric flux drops sharply at 4$\,\mu m$. Based on those facts we consider it likely that the source harbours a disk. It is the only one where our classification disagrees with \citet{luhman16}.

LRL40142 has slightly elevated flux levels at 3.6 and 4.5$\,\mu m$, compared to the 1900\,K template. Choosing a template with a slightly higher or lower temperature does not change the basic outcome. The SED peaks in the K-band, and fluxes drop towards longer wavelengths. There are no useful fluxes from C2D at wavelengths $>5\,\mu m$. Given that the SED declines it is reasonable to assume that the object does not have a disk.

LRLL6005 and LRL5209 have infrared fluxes that are largely consistent with the photosphere, within the considerable margin of error. Their SEDs are declining beyond the K-band. There are no C2D datapoints for longer wavelengths. In both cases we use the 1900\,K model as template, but the specific choice does not change the outcome. It is unlikely that any of these two has excess emission due to a disk based on the available data.

\begin{table*}
\caption{Planetary-mass objects in IC348 with potential disk. JHK photometry from UKIDSS (except J-band from \citet{preibisch03} for LRL30057). Spectral type and $A_J$ from \citet{luhman16}. IRAC photometry from this paper. Teff is the effective temperature of the model spectrum used for comparison in Figure \ref{fig:sed}. 'C2D' indicates if there are measurements from C2D at 5.8 and/or 8.0$\,\mu m$. The column 'disk' indicates whether or not there is evidence for a disk.}
\label{tab:pmo}
\begin{tabular*}{\textwidth}{@{\extracolsep{\fill}}lllcccccccll}
\hline
UGCS           & LRL   & SpT   & $A_J$ & J & H & K & I1 & I2 & Teff & C2D & disk \\
\hline
J034350.58+320317.5 & 1843  & M9    & 2.17 & 17.73  & 16.30 & 15.07 & 13.89 & 13.34 & 2200 & yes & yes \\ 
J034416.18+320541.0 & 4044  & M9    & 0.47 & 17.66  & 16.79 & 16.07 & 15.20 & 14.75 & 2100 & yes & yes\\  
J034451.99+315921.6 & 1379  & M9.7  & 0.29 & 17.61  & 16.68 & 15.85 & 14.95 & 14.56 & 2000 & yes & yes\\  
J034427.47+320624.2 & 30057 & L1    & 0.49 & 20.05  & 19.25 & 18.34 & 16.72 & 16.17 & 1900 & no  & yes\\  
J034449.33+320949.4 & 40023 & M9.5  & 0.39 & 18.66  & 17.66 & 16.89 & 15.76 & 15.56 & 1900 & no  & yes\\  
J034450.24+320635.5 &  5231 & M9.5  & 0.0  & 18.92  & 18.05 & 17.30 & 16.65 & 16.06 & 2000 & no  & yes\\  
J034400.75+320420.0 & 40142 & L0    & 0.0  & 19.34  & 18.42 & 17.60 & 16.76 & 16.51 & 1900 & no  & no?\\ 
J034425.92+320805.4 & 6005  & L1    & 0.49 & 19.29  & 18.30 & 17.46 & 16.66 & 16.17 & 1900 & no  & no?\\ 
J034434.54+320213.7 & 5209  & L1    & 0.49 & 19.43  & 18.16 & 17.50 & 16.63 & 16.31 & 1900 & no  & no?\\ 
\hline
\end{tabular*}
\end{table*}

\subsection{Disk fractions}

In this paper we have used the spectral type limit of M9 as a proxy for mass -- all 23 brown dwarfs with M9 or later in the \citet{luhman16} census are considered to be possible planetary-mass objects in IC348. For 13 of these we present new photometry from deep IRAC1 and IRAC2 images. For this subsample we test for the presence of a disk using the same consistent method. We find that six of those have a disk, corresponding to a disk fraction of 6/13 or 46\%. The uncertainty for this value was estimated by calculating how likely it is to observe 6 successes out of 13 trials, assuming a binomial distribution and a flat prior. This gives us a 1$\sigma$ confidence interval of 34\% to 59\%, which means the disk fraction is $46\pm ^{13}_{12}$\%.

\begin{figure}
\includegraphics[width=\columnwidth]{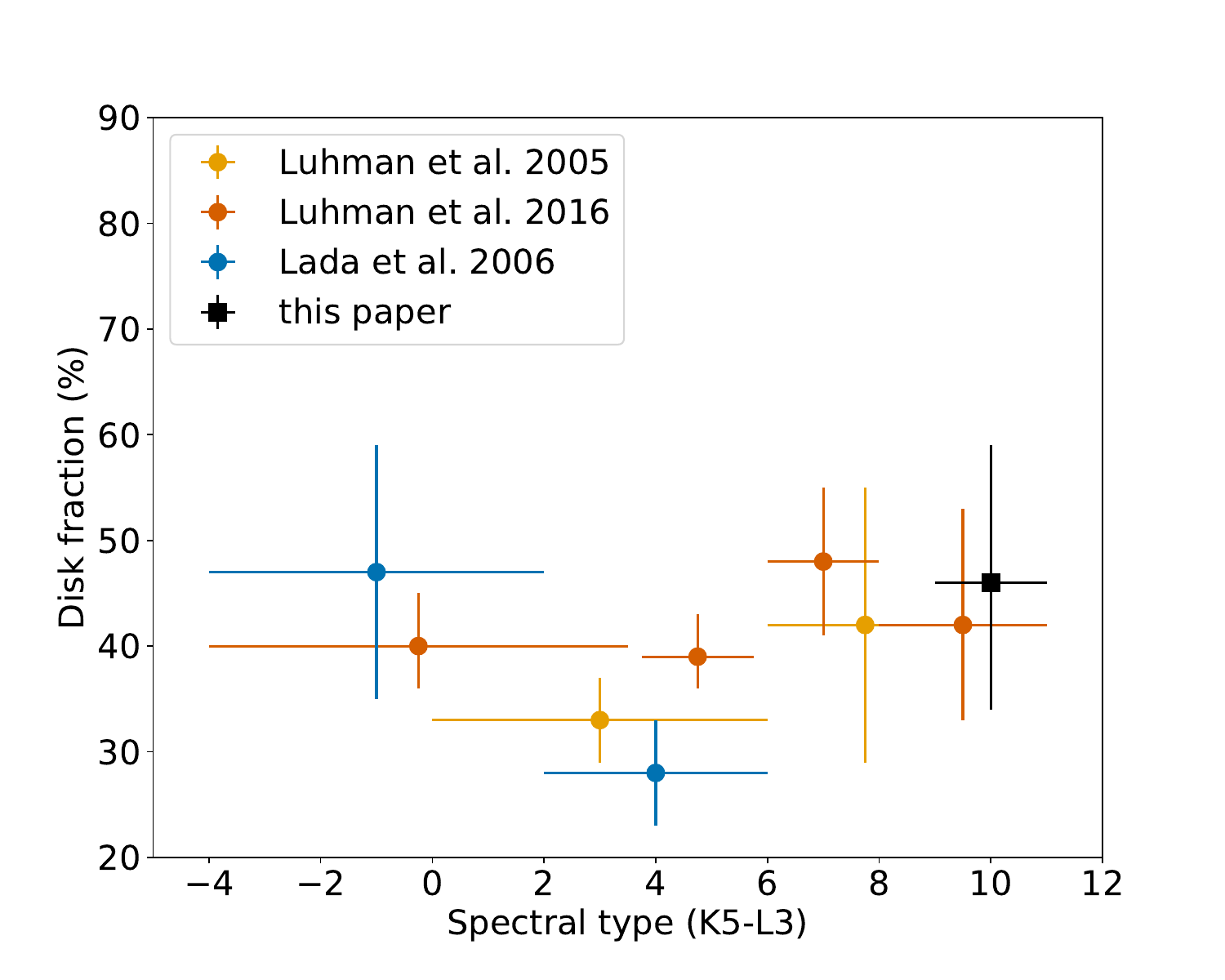}
\caption{Disk fraction in IC348 as a function of spectral type, including our new datapoints for planetary-mass objects. The data shown is from \citet{lada06,luhman05,luhman16}. For our estimate in black, we show the lower and upper limit.}
\label{fig:df1}
\end{figure}

In Figure \ref{fig:df1} we plot disk fractions for IC348 as a function of spectral types, including results from the literature combined with our new datapoint for planetary-mass sources. From \citet{lada06} we plot disk fractions of $47\pm 12$\% for spectral types K6-M2 and $28\pm 5$\% for M2-M6. Here we only consider the values for full disks, as anaemic disks, as classified by the authors, would be difficult to identify for faint late M/early L type objects. Also included are the values from \citet{luhman05} who measured $33\pm4$\% for M2-M6 and $42\pm13$\% for M6-M9. The \citet{luhman16} census gives disk fractions of 39\% for mid M, 48\% for M6-M8, and 42\% for later-type objects, with associated errorbars; these datapoints are also included in our figure. 

As seen in this figure, our datapoint is in agreement with all previous measurements of the disk fraction in the brown dwarf domain, within the (substantial) errorbars. It is very similar to the value for $>$M8 objects published in \citet{luhman16}. As already mentioned, our indicative estimate of $\sim 50$\% for M6-M9 objects is also in line with previous measurements.

The plot does show an overall trend of disk fraction with spectral type -- the disk fraction drops slightly from around 50\% to 30\% from mid K to early M spectral types (already noticed in the literature), but then rises slightly towards late M spectral types. An trend of increasing disk fraction with later spectral type has been noted for low-mass stars and brown dwarfs in the older Upper Scorpius star forming region \citep{luhman12}, but based on a small sample size. It may be an indication that disks among very low mass brown dwarfs live longer than in more massive objects.

\begin{figure}
\includegraphics[width=\columnwidth]{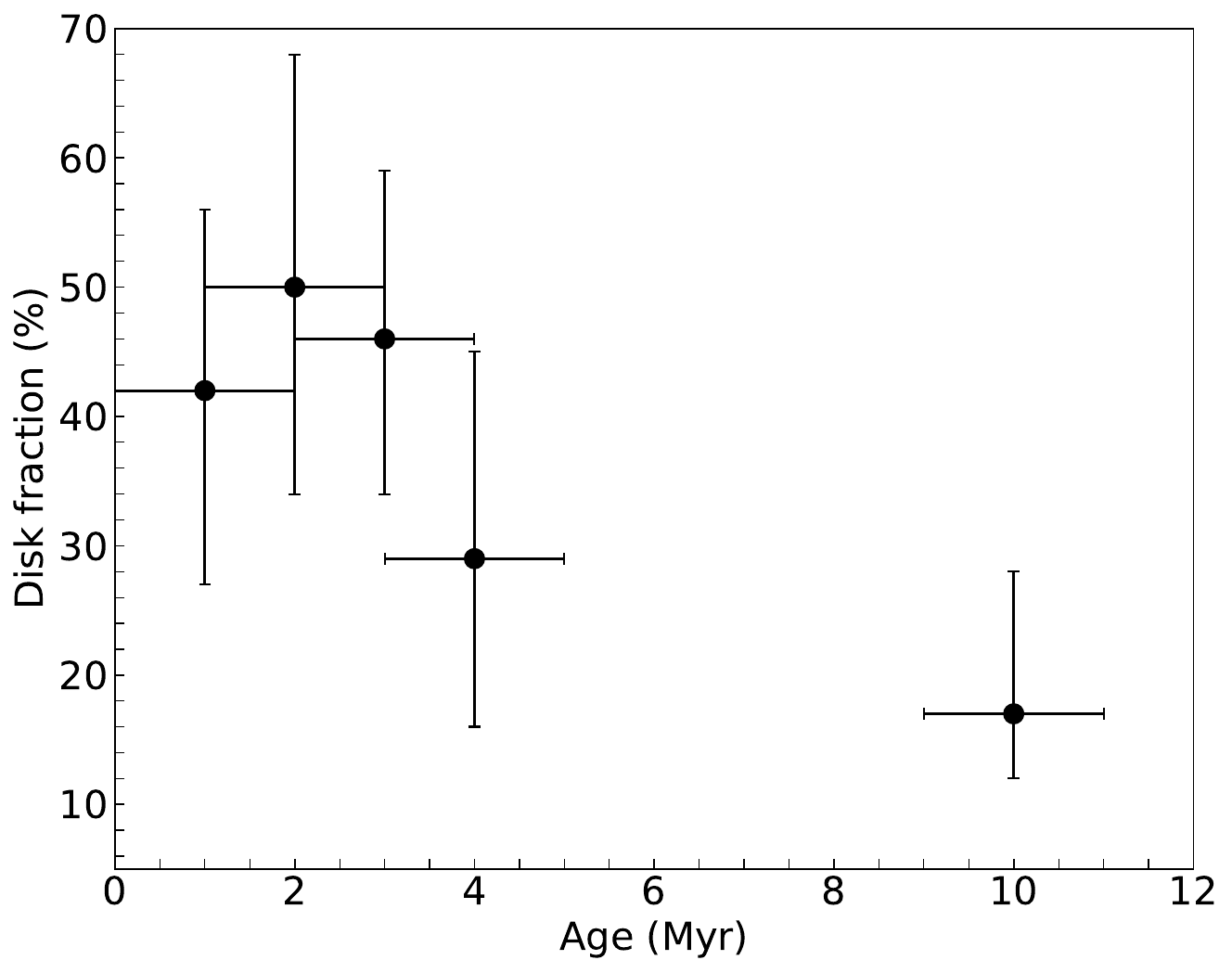}
\caption{Disk fraction among free-floating planetary-mass objects as a function of age. Included are datapoints for NGC1333 \citep[1 Myr][]{scholz23}, Chamaeleon-I \citep[2 Myr][]{luhman08}, IC348 (3 Myr, this paper), $\sigma$\,Orionis \citep[4 Myr][]{scholz08} and Upper Scorpius \citep[10\,Myr][]{luhman12}. Note that these are typical ages for these regions, the uncertainty is approximately $\pm 1$\,Myr.}
\label{fig:df2}
\end{figure}

With the new datapoint in IC348, we also have an opportunity to probe the evolution of disks around planetary-mass objects over time. In Figure \ref{fig:df2} we show disk fractions for five star forming regions, NGC1333 \citep{scholz23}, Chamaeleon-I \citep{luhman08}, IC348 (this paper), $\sigma$\,Orionis \citep{scholz08}, and Upper Scorpius \citep{luhman12}. The numbers are plotted at approximate ages, typical for ages cited in the literature, but the uncertainty (or spread) in age is at the minimum 1\,Myr. In all regions, the disk fractions have been determined for a spectral type range comparable to the one probed here (i.e. either M9 or later or M8 or later). Since disk fraction does not depend strongly on spectral type (see above), using those datapoints together is not introducing a bias. In all regions, the disk fractions are estimated at 3-5$\,\mu m$, i.e. the datapoints should be comparable. As show by this plot, disk fractions for planetary-mass objects stay in the 40-50\% range for at least 3-4\,Myr before they significantly drop off. Thus, the disk lifetime in planetary-mass objects is not strongly different from that in brown dwarfs or low-mass stars \citep{haisch01,manara23}. 

\section{Summary}
\label{sec:sum}

In this paper we search for disks around planetary-mass objects in the young cluster IC348. The study is based on deep Spitzer images, created by stacking 38 archival images taken for a time series study. The images cover 13 previously identified young brown dwarfs with spectral types of M9 to L3, corresponding to masses below or around the Deuterium burning limit. This is the core sample for this paper.

We measure infrared fluxes at 3.6 and 4.5$\,\mu m$ for our sample. Based on the $K-[4.5]\,\mu m$ colour, 9 of those 13 objects show signs of excess emission. We further examined these by comparing the full spectral energy distribution with model spectra. Six of them we classify as objects with disks, with multiple datapoints showing excess emission above the photosphere. The remaining three have SEDs largely consistent with a photosphere. The disk detections are excellent targets for future detailed studies of disks around planetary-mass objects.

The disk fraction in our sample is 6/13 or 46\%. This is entirely consistent with previous estimates in the literature for a similar spectral type range. Planetary-mass objects at ages of a few Myr have disk fractions comparable to higher mass brown dwarfs and low-mass stars. About a third to half of them retain their disks for several million years. The long-lived disks identified here may signal that there might be planets around objects which themselves have masses comparable to giant planets.

\section*{Acknowledgements}

We thank the anonymous referee for a constructive review that helped to improve this paper.
AS acknowledges support from the UKRI Science and Technology Facilities Council through grant ST/Y001419/1/. 
This research has made use of the VizieR catalogue access tool, CDS, Strasbourg, France (DOI: 10.26093/cds/vizier). The original description of the VizieR service was published in 2000, A\&AS 143, 23.

\section*{Data availability}

The imaging data underlying this article were accessed from the Spitzer Heritage Archive, at {\url https://irsa.ipac.caltech.edu/applications/Spitzer/SHA/}. In addition, data from peer-reviewed papers is used, all cited in this paper. The data for the colour-magnitude plots in Figure \ref{fig:kirac} is included in Table \ref{tab:phot}. The data for the spectral energy distributions in Figure \ref{fig:sed} is contained in Table \ref{tab:pmo}.


\bsp	
\label{lastpage}
\end{document}